    \documentstyle[epsf]{article} 
\makeatletter
    \def\section{\@startsection{section}{1}{\z@}%
    {-3.5ex plus -1ex minus -.5ex}{1.5ex plus.3ex}{\bf }}
    \def\subsection{\@startsection{subsection}{1}{\z@}%
    {-3.5ex plus-1ex minus-.5ex}{1.5ex plus.3ex}{\bf }} 
\makeatother
    
\begin{document}
    \hfill\parbox{4.77cm}{\Large\centering Annalen\\der
    Physik\\[-.2\baselineskip] {\small \underline{\copyright\ Johann
    Ambrosius Barth 1998}}} \vspace{.75cm}\newline{\Large\bf
Critical level statistics of a quantum Hall system 
\\[.2\baselineskip]
with Dirichlet boundary conditions
    }\vspace{.4cm}\newline{\bf   
H. Potempa and L. Schweitzer
    }\vspace{.4cm}\newline\small
Physikalisch-Technische Bundesanstalt, 
Bundesallee 100, D-38116 Braunschweig, Germany
    \vspace{.2cm}\newline 
Received 6. October 1998 
    \vspace{.4cm}\newline\begin{minipage}[h]{\textwidth}\baselineskip=10pt
    {\bf  Abstract.}
We investigate numerically the influence of Dirichlet boundary 
conditions on the nearest neighbor level spacing distribution $P(s)$
of a two-dimensional disordered tight-binding model in the presence 
of a strong perpendicular magnetic field.
From the calculation of the second moment of $P(s)$ it is shown that 
for Dirichlet boundary conditions, due to the presence of edge states, 
the position of the critical energy shifts with increasing system size
to the location of the critical energy for periodic boundary conditions. 
An extrapolation to infinite system size results in different critical 
(scale independent) $P(s)$ distributions for periodic and Dirichlet boundary 
conditions.
    \end{minipage}\vspace{.4cm} \newline {\bf  Keywords:}
QHE, level statistics, critical spacing distribution, boundary conditions
    \newline\vspace{.2cm} \normalsize

\section{Introduction} 
In recent years the statistics of energy eigenvalues has become an important 
tool for investigating the localization properties of disordered electronic 
systems \cite{Efe83,AS86}. 
In most cases the Anderson model or one of its variants were used to describe 
the behavior of non-interacting electrons in three-dimensional disordered 
systems which possess orthogonal 
\cite{AZKS88,Sea93,HS93a,HoS94a,Eva94,ZK95a,ZK95b,ZK97}, 
unitary \cite{HoS94b,BSZK96,DBZK98}, and symplectic \cite{KOSO96} symmetry.
Also two-dimensional systems with orthogonal \cite{ZBK96,Cea98}, 
symplectic \cite{SZ95,Eva95,SZ97}, and unitary (random flux) \cite{BSK98} 
symmetry have been studied. Even in the quantum Hall case (strong magnetic 
field) \cite{HS92,OO95,FAB95,BS97} level statistics was used to examine
the localization properties. 
  
It has been shown that in the limit of infinite system size, $L\to\infty$,
the statistics of 
the uncorrelated eigenvalues in the insulating regime is specified by the 
Poisson form whereas in the diffusive regime the correlations of energy 
eigenvalues are well described \cite{Efe83} by random matrix theory (RMT), 
see e.\,g., \cite{Meh91, GMW98}. 
There are, however, also known deviations \cite{AS86,KM94,BM95}
from the universal RMT results.  
These deviations arise when the mean level spacing $\Delta$ is equal or larger 
than the Thouless energy $E_c=\hbar D/L^2$ which is the inverse of the 
classical diffusion time.
The discrepancy is not really astonishing because the respective Hamilton 
matrix for the Anderson model with its large number of zeros is quite 
different from the random matrices considered in RMT 
\cite{Wig51, Dys62, Meh91}. 

Nevertheless, the success of extracting the critical properties that govern 
the localiza\-tion-delocalization transition directly from level statistics 
\cite{Sea93,HoS94a,BM95,BS97,SZ97} has stimulated  detailed 
investigations of the new class of critical level statistics. 
In contrast to the metallic and insulating phases where the level statistics
assume their respective universal form in the limit of infinite system size, 
the critical P(s) was found to be scale invariant. 

In particular, results on the energy spacing distribution of successive 
eigenvalues, $P(s)$, where $s=|E_{i+1}-E_i|/\Delta$,
have been frequently reported in the literature.
Until recently the form of the critical $P(s)$ was known only from numerical
studies. However, the promising recent progress of analytical theories 
\cite{AKL95,CLS96,CKL96,Can96,Nis99} will help to understand the critical
eigenvalue correlations and to clarify the supposed relation \cite{AZKS88} 
between the large-$s$ decay of $P(s)$ and the level number variance, 
$\Sigma^2(\langle N\rangle)=\chi \langle N\rangle$, 
which was shown to be connected 
with the multifractal exponent $D(2)$ of the critical eigenstates via 
$\chi=d-D(2)/(2d)$ \cite{CKL96}.

Even in two-dimensional quantum Hall effect (QHE) systems, where the 
localization length diverges at singular energies in the center of the 
Landau bands and no complete localization-delocalization transition exists 
(absence of an energy range of extended states), critical level statistics 
can still be observed 
\cite{HS92,OO95,FAB95,BS97}.

Recently, the eigenvalue statistics directly at the transition was 
shown to depend on the boundary conditions in 3d orthogonal \cite{BMP98} 
and 2d symplectic systems \cite{SP98}, and also on the 
shape of the 3d sample \cite{PS98}. For a QHE system, however, it has been
suggested \cite{EK98} that the dependence of the level statistics on the 
boundary conditions is absent when an appropriate shift of the critical 
energy is taken into account. In the following we address the question, 
whether the critical 
level statistics of the QHE system behaves differently to a change of the 
boundary conditions.

\section{QHE-model and numerical method}

The two-dimensional system of non-interacting electrons in the presence of 
impurity scattering and a strong perpendicular magnetic field can be described 
\cite{SKM84a} by a tight-binding Hamiltonian on a square lattice with diagonal 
disorder. 
The magnetic field enters via complex phases in the transfer terms that 
cause the electronic motion within the $xy$-plane. In the Landau gauge the 
vector potential is taken as $A=(0,Bx,0)$ which results in a magnetic flux
density in the $z$-direction.
The Hamilton matrix is
\begin{eqnarray}
\lefteqn{(H\psi)(x,y)=\epsilon(x,y)\,\psi(x,y)+V\,[\psi(x+a,y)+\psi(x-a,y)+}
\nonumber\\
& & \exp(-i2\pi\alpha_Bx/a)\,\psi(x,y+a)+
\exp(i2\pi\alpha_Bx/a)\,\psi(x,y-a)],
\end{eqnarray}
where $a=1$ and $V=1$ are taken as the unit of length and energy, 
respectively. The magnetic field $B$ is chosen to be commensurate with the
lattice and $\alpha_B=a^2eB/h$ denotes the number of flux quanta per 
plaquette. The boundary conditions are changed from periodic (PBC) to
Dirichlet (DBC) by setting $V=0$ along the edges. The diagonal disorder
potentials $\epsilon(x,y)$ are a set of independent random numbers drawn 
from an interval, $-W/2 \leq \epsilon \leq W/2$, with constant 
probability density $1/W$.

Square systems of size $L/a=64$, 96, 128, 256  are considered
with disorder strength $W/V=2.5$. Taking $\alpha_B=1/8$, the ratio of 
system size $L$ to the magnetic length $l=(\hbar/(eB))^{1/2}$ is about 201
for the largest samples.
The corresponding large sparse hermitian matrices were directly diagonalized 
by means of a Lanczos algorithm based on \cite{CW85} which was adapted to run 
efficiently on DEC-Alpha workstations. 
Many realizations of the disorder potentials were calculated so that 
the number of accumulated eigenvalues within a narrow interval around the
critical energy of the lowest Landau band exceeded $2\cdot 10^5$ even for
$L/a=256$. A spectral unfolding procedure was applied in order to eliminate 
global changes in the density of states.

\begin{figure}
\epsfxsize12.cm\epsfbox{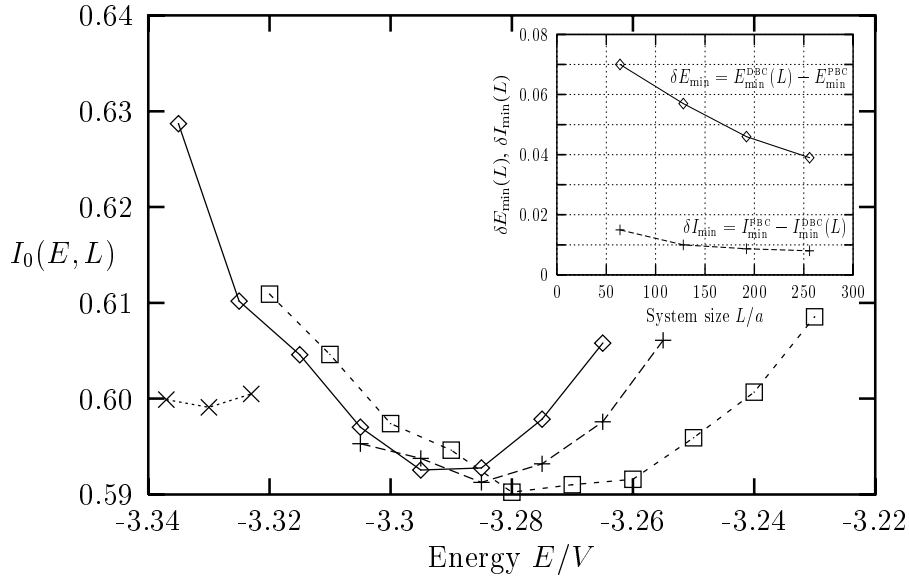}
\caption[]{\small The energy and size dependence of $I_0$ for DBC: ($\Box$) 
$L/a=128$, ($+$) $L/a=192$, ($\Diamond$) $L/a=256$. In contrast, for PBC the 
energetic position of the $I_0$-minimum is size independent: ($\times$) 
$L/a=128$. The lines only connect points that
belong together. The inset shows the size dependent shift of the energetic 
position and the value of the $I_0$-minima. The minimum value of $I_0$ seems 
to saturate at about $0.592$.\label{I0_E_L}}
\end{figure}

\section{Results and discussion}
As mentioned in the introduction, the critical disorder $W_c$ which defines
the position of the divergence of the correlation length, 
$\xi(W)\sim |W-W_c|^{-\nu}$, does not change significantly in 3d orthogonal 
\cite{BMP98} and 2d symplectic \cite{SP98} systems when the periodic boundary 
conditions (PBC) are replaced by Dirichlet boundary conditions (DBC).
At $W_c$ and for finite systems, there exists a broad energy interval for 
which the localization length exceeds the system size $L$ so that these states
show critical behavior.  
For the 2d QHE-System, on the other hand, the divergence of the localization 
length, $\lambda\sim |E-E_n|^{-\nu}$, takes place at the positions of the 
critical energies, $E_n(W)$, which depend on the disorder strength. 
Therefore, only a small critical energy window around $E_n$ exists where 
$\lambda(E) >> L$.

To determine the position of the critical energy in the lowest Landau band, 
$E_0$, we have calculated the energy dependence of the second moment 
$I_0(E,L)$ 
of the level spacing distribution, $I_0(E,L)=1/2\int_0^{\infty} s^2 P(s)\,ds$,
for PBC and DBC. The minimum of $I_0$ indicates the position of $E_0$ which
is shown in Fig.~\ref{I0_E_L} for disorder $W/V=2.5$. While the energetic
position of the minimum of $I_0$ does not change with the system size $L$ 
for PBC, a pronounced shift is visible in the case of DBC. 
The reason for the shift originates in the edge states which are 
introduced by the application of DBC. However, their influence on the
level statistics vanishes with increasing system size. 
This is shown in the inset of Fig.~\ref{I0_E_L} where the difference
$\delta E_{\rm min}=E_{\min}^{\rm\tiny DBC}(L)-E_{\min}^{\rm\tiny PBC}$ 
between the energetic positions of the minimum of $I_0$ for DBC and PBC is 
plotted. 
There, also the difference 
$\delta I_{\rm min}=I_{\rm min}^{\rm PBC}-I_{\rm min}^{\rm DBC}(L)$ 
in the corresponding values of $I_0$ is shown. While the energetic position
of $I_{\rm min}^{\rm DBC }$ moves towards $E/V=-3.33$ found for PBC, the
absolute value seems to saturate at $I_{\rm min}^{\rm DBC}=0.592$ which
is smaller than the calculated $I_{\rm min}^{\rm PBC}=0.6$.   

\begin{figure}
\epsfxsize12.cm\epsfbox{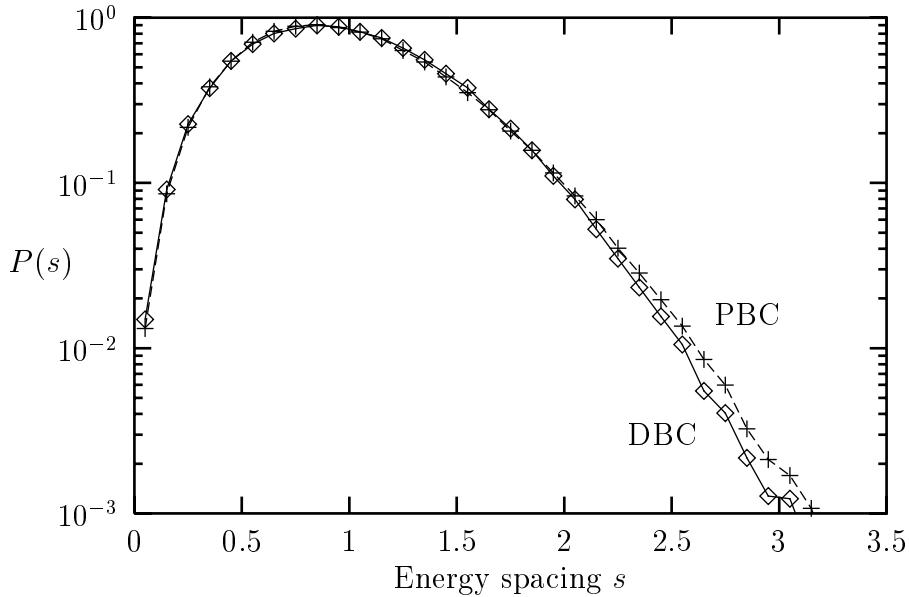}
\caption[]{\small The critical (scale independent) energy spacing 
distributions $P(s)$ for periodic boundary conditions (PBC, $+$) and 
Dirichlet boundary conditions (DBC, $\Diamond$) versus energy spacing $s$ in 
the lowest Landau band.\label{pofs}}
\end{figure}

Now, we address the question, whether the critical level spacing distributions
differ for different boundary conditions. In Fig.~\ref{pofs} the  
semi-logarithmic plots of $P(s)$ are shown for PBC with system size 
$L/a=128$ and for DBC with $L/a=256$. A small, but significant 
difference is observed between the curves for PBC and DBC for
spacings $s > 2$. In contrast to the 3d orthogonal and the 2d symplectic 
case the large-$s$ slope is steeper for DBC than for PBC in the QHE system. 
Within the numerical uncertainty of our data, no size dependence could 
be detected for $P(s)$ with PBC in the range
from $L/a=32$ to 128 and for DBC in the range $L/a=128$ up to 256. 
Therefore, we have 
to conclude that for the QHE system in the limit $L \to \infty$ the critical 
distributions are different for PBC and DBC, but the critical energies are
the same. The reason for this peculiar behavior presumably originates in the 
different topology of the systems \cite{KY98}. 
However, in order to detect the small difference in $P(s)$, strong finite 
size effects have to be overcome. The latter are due to the appearance of 
extended edge states that reside along the boundaries of the QHE system 
when DBC are applied.



    \end{document}